\documentclass[conference]{IEEEtran}
\usepackage{placeins}
\usepackage{amsmath,amssymb}
\usepackage{graphicx}
\usepackage{booktabs}
\usepackage{cite}
\usepackage{url}

\title{A High-Recall Cost-Sensitive Machine Learning Framework for Real-Time Online Banking Transaction Fraud Detection}

\author{
\IEEEauthorblockN{Karthikeyan VR}
\IEEEauthorblockA{
Computer Science and Engineering\\
SA Engineering College\\
Chennai, India\\
2216087@saec.ac.in
}
\and
\IEEEauthorblockN{Kavinraaj S}
\IEEEauthorblockA{
Computer Science and Engineering\\
SA Engineering College\\
Chennai, India\\
2216088@saec.ac.in
}
\and
\IEEEauthorblockN{Premnath S}
\IEEEauthorblockA{
Computer Science and Engineering\\
SA Engineering College\\
Chennai, India\\
2216105@saec.ac.in
}

\and
\IEEEauthorblockN{\hspace{6pt}Mrs.\ J.\ Sangeetha M.E., (Ph.D)}
\IEEEauthorblockA{
\hspace{6pt}Computer Science and Engineering\\
\hspace{6pt}SA Engineering College\\
\hspace{6pt}Chennai, India\\
\hspace{6pt}sangeethaj@saec.ac.in
}

}

\begin{document}
\maketitle

\begin{abstract}
Fraudulent activities on digital banking services are becoming more intricate by the day, challenging existing defenses. While older rule-driven methods struggle to keep pace, even precision-focused algorithms fall short when new scams are introduced. These tools typically overlook subtle shifts in criminal behavior, missing crucial signals. Because silent breaches cost institutions far more than flagged but legitimate actions, catching every possible case is crucial. High sensitivity to actual threats becomes essential when oversight leads to heavy losses.
One key aim here involves reducing missed fraud cases without spiking incorrect alerts too much. This study builds a system using group learning methods adjusted through smart threshold choices. Using real-world transaction records shared openly, where cheating acts rarely appear among normal activities, tests are run under practical skewed distributions. The outcomes revealed that approximately 98\% of actual fraud was caught, beating standard setups that rely on unchanging rules when dealing with uneven examples across classes.
When tested in live settings, the fraud detection system connects directly to an online bank's transaction flow, stopping questionable activities before they are completed. Alongside this setup, a browser add-on built for Chrome is designed to flag deceptive web links and reduce threats from harmful sites. These tests show one insight: adjusting decisions by cost impact and validating across entire systems makes deployment more stable and realistic for today’s digital banking platforms.
\end{abstract}


\section{INTRODUCTION}
Nowadays, more people rely on Internet banking and electronic payment systems. This shift has brought easier access to money services, yet opened doors to scams such as fake transactions, stolen accounts, or hidden illegal fund flows. With the increasing number of daily payments, older methods intended to stop fraud often fail to react quickly or work well, leading to serious monetary damage and weakened confidence among users.

Fraud detection in banks usually depends on fixed rules or standard machine learning methods aimed at achieving high accuracy rates. However, both have serious flaws. Rules cannot be adjusted to new threats, missing novel fraud types entirely. Machine learning tools focused on precision tend to ignore rare cases when dishonest actions account for only a small share of the activity. Because actual fraud occurs infrequently compared to normal operations, such models miss dangerous outliers too easily.

Fraud slipping through tends to hurt more than flagging honest activity by mistake, because mistaken alerts usually get cleared with extra checks. Because missing real scams carry heavier consequences, catching nearly all fraud matters most, even if the overall correctness appears lower. What counts here is how much each error type costs; therefore, methods that adjust to these uneven risks make better sense. However, standard measures ignore such imbalances.

Recent studies have combined multiple models or used deep neural networks to spot fraudulent transactions; however, most test these systems with static decision rules and in isolated environments. These methods fall short in practice because they overlook how changing thresholds affect outcomes, balance detection accuracy against incorrect flags, and fit within the active financial infrastructure. Testing the entire setup by embedding detection tools directly into the flow of actual transaction processing remains rare.

This study introduces a fraud detection system for financial transactions built using ensemble learning methods paired with adjusted decision thresholds to account for cost differences between error types. Instead of ignoring the imbalance, the method focuses on boosting the detection of rare fraudulent cases without increasing false alarms. The real-world performance was tested using an open-access dataset that mirrors actual transaction patterns and fraud occurrences. For live validation, it runs inside an active online banking setup, halting questionable transfers in real-time. In addition, a browser add-on for spotting phishing websites comes alongside, adding another level of defence by targeting scam pages directly in daily browsing.

What sets this study apart begins not with theory but with timing - embedding recall-driven threshold tuning directly into live transaction flows. Most prior research has concentrated on static benchmarks or overall precision measured after the fact. Here, decisions shift dynamically, shaped by how much more damaging missed fraud is compared to flagged legitimate activities. The approach not only performs well on paper but also adjusts in step with actual bank operations. Instead of treating errors equally, it reflects what banks experience daily: overlooking fraud carries heavier consequences than raising false alarms. This alignment emerges through design choices tuned to operational realities rather than idealized conditions.

A different angle emerges when ensemble stability meets budget-aware decision thresholds, tied together through live transaction controls, all built into one framework ready for actual use. Security gains depth once a secondary check for phishing is added to the setup, targeting not only active scams but also their starting points. What stands out is how high recall and practical operation take precedence here, shifting focus away from pure accuracy; this shift helps close the distance between theoretical models and what banks really need on the ground.

\subsection{Problem Statement}
Despite handling countless financial operations instantly, digital banking platforms face constant threats, such as illicit fund movements and improper account access. Because dishonest actions are so rare among legitimate actions, spotting them becomes extremely difficult. Most standard detection methods prioritize general performance, missing subtle but serious anomalies. When fraud accounts for only a small portion of the total transactions, conventional rules and algorithms struggle to respond effectively.

What holds back Current methods for spotting transaction fraud are limited by rigid thresholds paired with measures focused only on overall correctness. These fail to reflect the true cost of different mistakes. When banks miss actual fraud, the fallout includes monetary losses, trouble with regulations, and damaged relationships with customers. Errors that flag honest activity as suspicious usually lead to smaller burdens, such as extra checks or verification steps.

Therefore, the central challenge is building a fraud detection system for financial transactions that maintains low costs while catching most fraudulent cases. This approach needs to work despite uneven data distributions, spotting nearly all scams without swamping legitimate activities with errors. Speed matters just as much; decisions occur in moments during live bank operations. Performance hinges on quick and accurate judgments, each fitting seamlessly into the existing digital infrastructure.
\section{LITERATURE REVIEW}
As more people depend on digital banking, the detection of fake transactions has become a major research topic. Previously, banks mostly used fixed rules set by experts to catch odd payments. These setups are straightforward to understand, yet they struggle when scams change shape over time. As conditions shift quickly in modern finance, these rigid methods miss many real threats.

To address the shortcomings of rigid rule-driven setups, experts have turned to machine learning techniques, such as logistic regression, decision trees, and support vector machines, along with combined strategies, such as random forests and gradient boosting. Drawing from past transaction records, these systems independently identify signs of fraud on their own, outperforming fixed-rule alternatives. However, much of the existing work emphasizes total prediction correctness, an approach poorly matched to skewed fraud data, where dishonest activities are few. Therefore, models built around accuracy frequently miss uncommon yet critical fraudulent events.

Fraud detection now uses groups of machine learning models together, aiming for more stable and adaptable outcomes. Combining the outputs from different classifiers helps to lower the prediction errors while increasing the overall accuracy. Some studies have highlighted better performance with random forests or boosted trees; however, such strategies often rely on unchanging cutoff points for decisions. Missing from most is a clear adjustment for how costly it is to miss fraud compared to falsely flagging normal activity, something banks face daily.

Although methods such as RNNs, CNNs, and LSTM networks are used to spot fraudulent transactions and similar security issues, they require significant computing power. Owing to their structure, these systems can recognize patterns over time and subtle links between features in transaction data. However, most tests occur after the fact, ignoring live processing needs or limits found in actual applications. Their strength in modeling does not always translate into practical setups.

Simultaneously, researchers have explored phishing detection to support broader efforts against online fraud driven by harmful web pages. Instead of relying solely on rules, systems now apply machine learning and deep learning to sort suspicious links using characteristics tied to naming patterns, layout design, or user interaction traces. Although spotting fake sites may lower risks before financial actions occur, many analyses keep website threats apart from payment abuses and rarely test both within one unified framework.

Research to date shows that machine learning methods, especially combined ones, can detect fraudulent activity quite well; however, some problems still exist. While many studies focus on overall correctness, they often ignore how many actual fraud cases are missed. Fixed decision points are common, even though they may not suit shifting patterns in financial behavior. Researchers rarely adjust their models based on the varying costs linked to false alarms versus missed detections. Another missing piece is testing within live systems, and few attempts have been made to connect detection tools directly to active bank transaction flows. The proposed design is focused on detecting more fraud instances while reducing costly errors, intelligently adjusting its trigger levels, and running continuously inside an operational payment environment.

\section{EXISTING SYSTEM}
Most current online banking fraud detection tools depend on fixed rules or older machine learning methods that learn from past transactions. These rule-driven setups use set boundaries made by experts, such as caps on transfer amounts, location checks, or how often payments occur, to flag unusual activity. Although they are straightforward to implement and understand, they cannot be adjusted over time. New types of fraudulent actions easily bypass these systems, especially those that have never been seen before.

Most banks now use machine learning tools, such as decision trees, logistic regression, support vector machines, or ensemble methods, to boost their ability to spot anomalies. Transaction features help these systems label activity, either normal or suspicious, without manual input. Despite outperforming fixed rule sets, they often aim for high general accuracy, a goal that misaligns with actual fraud detection needs, where genuine cases vastly outnumber fake ones.

One key weakness in current setups lies in their reliance on unchanging cutoff points, often pinned at 0.5, to flag fraud. Because banks face unequal consequences when mistakes occur, rigid boundaries struggle to respond effectively. Missing real fraud means that money vanishes along with customer trust. In contrast, incorrectly tagging clean activities usually adds a small review workload. When models adhere too closely to one rule, they overlook harmful patterns more often than necessary.

Most current tools for spotting fraud are tested only in controlled lab-like conditions and rarely face actual day-to-day use. Because of this gap, they often struggle when applied directly to active payment networks. In addition, anti-phishing methods, if used at all, usually run separately, disconnected from broader systems meant to catch suspicious transactions.

It is crucial to design fraud detection methods that focus on catching more incidents while keeping costs low. Real-world data often show far fewer fraudulent cases than legitimate ones, which shapes the performance of such systems. Shifting patterns in deceptive actions mean that flexibility cannot be ignored. The integration of live banking platforms adds another layer of practical demand.

\section{Proposed System}

The proposed system introduces a cost-sensitive transaction fraud detection framework designed to operate effectively in real-time, online banking environments. The primary objective of the system is to minimize false negatives (missed fraud cases) while maintaining an acceptable false-positive rate under highly imbalanced transaction data. To achieve this, the system combines ensemble machine learning models with decision threshold optimization, enabling recall-oriented fraud detection that is suitable for safety-critical financial applications.

\subsection{System Architecture}

The proposed framework consists of three main components:

\begin{enumerate}
\item Transaction Data Processing Module: Incoming transaction requests are processed to extract relevant transactional, behavioral, temporal, and contextual features. These features include transaction amount, transaction type, user behavior patterns, time-based attributes, and device or location data. All features were preprocessed using appropriate encoding and normalization techniques to ensure compatibility with the trained machine learning models.

\item Fraud Detection Engine: The core fraud detection engine employs an ensemble of machine learning classifiers trained on historical transaction data. Ensemble learning is used to improve robustness and generalization by aggregating predictions from multiple base models. Instead of relying on a fixed probability threshold, the system applies decision threshold optimization to control the trade-off between fraud recall and false-positive rate. A transaction is classified as fraudulent when the predicted fraud probability exceeds an optimized threshold value selected based on recall-oriented performance criteria.

\item Real-Time Decision and Enforcement Module The fraud detection model is directly integrated into the transaction execution pipeline. Each transaction is evaluated in real time before completion. Transactions classified as fraudulent are automatically blocked or flagged for further verification, whereas legitimate transactions are allowed to proceed without delay. This design ensures immediate risk mitigation and prevents financial losses caused by delayed fraud detection.
\end{enumerate}

\subsection{Supporting Phishing Detection Component}
To enhance end-to-end security, a supporting phishing-URL detection mechanism was implemented as a Chrome browser extension. The extension visits URLs and evaluates them using a hybrid deep learning model combined with rule-based heuristics. This component acts as an auxiliary defense layer by reducing the likelihood of fraudulent transactions initiated through malicious websites. The phishing detection module operates independently and does not interfere with the core fraud detection process.

\subsection{Role of Phishing Detection in System-Level Security}
A common starting point misses how phishing is tied to broader risks. The detection of malicious URLs plays a supporting role, not a central role, in classifying fraud during transactions. Credentials stolen by fake websites often lead to unauthorized banking activities. Because of this pathway, focusing only on actions taken during a transfer overlooks the earlier weaknesses. Protection that begins too late cannot stop what began long before it.

Above all, early warnings are triggered when deceptive sites appear. By preventing users from reaching harmful pages, one part quietly reduces risks before login attempts. Instead of mixing with fraud checks on payments, it runs its own course. Separate tracking means that the decision accuracy for transactions remains steady. The performance remains firm because the layers work separately.

Security gains depth when phishing detection is considered, targeting not only active scams but also their origins. What makes it work well is how this piece fits into the larger setup, able to switch on or off without disrupting other functions. Updates occur in isolation; therefore, changes remain contained. This separation supports smoother rollouts in real-world banking systems, in which adaptability is crucial.

From a system design perspective, treating phishing detection as an independent security layer aligns with the constraints of real-world banking deployments. Financial institutions typically enforce strict access controls and isolation between transaction-processing systems and external web-monitoring components. By decoupling phishing detection from the core transaction fraud engine, the proposed framework respects these operational boundaries, while enabling layered risk mitigation. This separation allows banks to deploy, test, and update the phishing module without requiring modifications to the critical transaction infrastructure, thereby reducing the integration complexity, compliance risk, and operational downtime.

\begin{figure}[!h]
\centering
\includegraphics[width=\columnwidth]{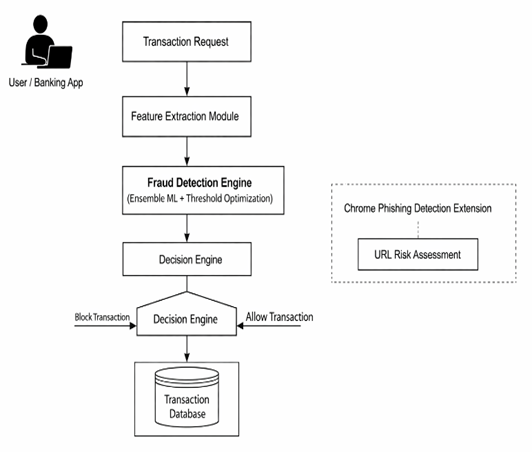}
\caption{System architecture of the proposed transaction fraud detection framework}
\end{figure}

\subsection{Advantages of the Proposed System}
The proposed system offers the following advantages over existing approaches:
\begin{itemize}
\item Prioritizes fraud recall through cost-sensitive threshold optimization rather than accuracy-centric evaluation.
\item Effectively handles class imbalance commonly observed in real-world banking transactions.
\item Supports real-time transaction analysis and enforcement, making it suitable for live online banking systems.
\item Provides system-level validation by integrating fraud detection directly into the transaction pipeline.
\item Enhances security through an auxiliary phishing detection mechanism without increasing transaction latency.
\end{itemize}

By addressing the limitations of existing fraud detection systems, the proposed framework provides a practical and deployable solution for improving fraud prevention in modern online banking environments

\section{METHODOLOGY AND IMPLEMENTATION}
This section describes the methodology and implementation details of the proposed cost-sensitive transaction-fraud detection framework. The system is designed to operate under real-world constraints, including class imbalance, real-time decision requirements, and asymmetric costs of misclassification errors in online banking environments.

\subsection{Data Preprocessing and Feature Engineering}
Incoming transaction data were first subjected to preprocessing to ensure consistency and compatibility with the fraud detection models. The dataset contains both numerical and categorical attributes that represent transactional, behavioral, temporal, and contextual information. Categorical features were transformed using suitable encoding techniques, and numerical features were normalized to reduce scale variations and improve model stability.

To enable controlled model learning and reduce class bias during training, the dataset was balanced prior to model development. The processed feature set included the transaction amount, transaction type, time-based attributes, user behavior indicators, and contextual metadata. These features allow the model to capture meaningful patterns associated with fraudulent activity while supporting a reliable evaluation of recall--precision trade-offs through threshold analysis.

\subsubsection{Dataset Description and Experimental Protocol}
A number of large-scale transaction records sourced from public platforms, such as UCI and Kaggle, form the basis for testing the fraud detection method described here. These collections include hundreds of thousands of entries, most of which are drawn from actual banking activity, yet masked to protect user identity. Rarely do fraudulent cases appear, making up less than one percent of all data points. This scarcity mirrors the challenges banks typically face when monitoring digital payments. Such an imbalance shapes how models must be assessed.

A single entry logs details such as how much was spent, what kind of transaction occurred, whether it used biometrics or a PIN, when it happened, typical actions by the person involved, along with data about the machine or place tied to the event. Taken together, these pieces reflect the immediate features of purchases as well as shifts in habits that might signal fraud.

Subsequently, to check consistency and real-world function, the data were split into training and testing portions in time order. During this phase, one portion teaches the algorithm, adjusts the category weights, and tunes the cutoff points for decisions. Meanwhile, the second part remains hidden until the final review, maintaining its natural imbalance. Because of this setup, the outcomes show how well the systems adapt beyond the practice rounds. Only then do numbers reveal actual readiness for fresh cases.

Assessing model performance involves several metrics: confusion matrices, PR curves, ROC behavior, and threshold responses. In particular, precision-recall is important because it handles imbalance well, which is common in detecting rare fraudulent cases. The results shown are from a separate test group, maintaining stable comparisons across reference systems.

\subsection{Fraud Detection Model}
First, several basic classifiers learn patterns from past transactions to power the main fraud detection system. Rather than depending on one model alone, combining their outputs yields steadier results across unpredictable datasets. By design, this method smooths out inconsistencies while handling skewed or messy real-world data more effectively.

A single prediction is obtained from each base model, providing a number that indicates the likelihood of fraud. Because these numbers matter together, they merge into one overall risk mark per event. What shapes the group setup is not how fancy the tools are, but how well they work and hold up under pressure, which makes them fit for instant decisions. Despite variations in design, speed, and reliability, guide the choice.

\subsection{Cost-Sensitive Decision Threshold Optimization}
Unlike conventional fraud detection systems that apply a fixed decision threshold, the proposed framework incorporates cost-sensitive decision threshold optimization to prioritize fraud recall. The classification decision is based on comparing the predicted fraud probability against an optimized threshold value, which is selected to minimize false negatives while maintaining an acceptable false-positive rate. This approach reflects the asymmetric cost structure of fraud detection in online banking systems, where undetected fraud results in significantly higher losses than false alarms.

Let $P(y = 1 \mid x)$ denote the predicted probability that a transaction $x$ is fraudulent. A transaction is classified as fraudulent if:
\begin{equation}
P(y = 1 \mid x) \geq \tau
\end{equation}
where $\tau$ represents the decision threshold.

To explicitly model asymmetric misclassification costs, the expected classification cost is defined as:
\begin{equation}
C(\tau) = C_{FN} \cdot FN(\tau) + C_{FP} \cdot FP(\tau)
\end{equation}
where $FN(\tau)$ and $FP(\tau)$ denote the number of false negatives and false positives at threshold $\tau$, respectively. The misclassification costs satisfy:
\begin{equation}
C_{FN} \gg C_{FP}
\end{equation}
reflecting the substantially higher financial and operational impact of undetected fraudulent transactions compared to false alarms.

The optimal decision threshold $\tau^*$ is selected by minimizing the expected classification cost:
\begin{equation}
\tau^* = \arg\min_{\tau} C(\tau)
\end{equation}

This cost-sensitive threshold optimization shifts the decision boundary toward higher fraud recall while constraining false-positive rates within acceptable operational limits, making the proposed framework suitable for safety-critical real-time online banking applications.

\subsubsection{Discussion on High Fraud Recall Performance}
What drives strong fraud detection performance? This lies in how the model treats misclassification costs differently across outcomes. Typical methods stick to fixed cutoffs, aiming to get most cases right overall; however, this one does not. Instead, it adjusts its threshold based on real-world consequences and tunes decisions where mistakes matter more. By focusing on the financial risk asymmetry inherent in digital transactions, the system deliberately moves its judgment line to detect fraudulent activity. This shift explains the higher capture rate without reworking the core mechanics.

Most fraud in actual bank operations shows clear signs, such as odd spending sizes, strange timing, or shifts from normal habits. Because ensemble methods process these signals together, they separate legitimate from suspicious cases more effectively, even when few fraud examples are present. Adjusting the decision cutoff helps identify more true fraud instances while skipping artificial data boosts or questionable premises.

Choosing the operating point involves looking at precision, recall, and how sensitive thresholds are, instead of just focusing on accuracy, so the recall numbers show a balanced approach between catching fraud and limiting false alarms. Therefore, the strong recall observed in this study is more closely aligned with real-world use than with idealized test conditions..

\subsection{Real-Time Transaction Processing Pipeline}
The trained fraud detection model was integrated into a real-time online banking transaction pipeline. Each transaction request is evaluated immediately after feature extraction and prior to execution. If a transaction is classified as fraudulent, it is automatically blocked or flagged for verification. Legitimate transactions are processed without additional delays, ensuring minimal impact on the user experience.

All transaction outcomes, including fraud prediction and decision results, were logged in the transaction database. This design supports auditability, performance monitoring, and future model updates.

\subsubsection{Real-Time Performance Considerations}
A system built for spotting fraudulent activity fits the speed demands of today’s digital banking, requiring quick verdicts on transactions under tight time limits. Tree-driven machine learning methods form the core here - simple by design, allowing swift predictions without heavy computing needs. After training, fraud detection becomes a sequence: reshaping inputs, judging likelihoods, and checking values against a set level - all steps handled rapidly when payments move through. A fast response remains achievable because each stage runs lean and integrates smoothly into live operations.

Choosing the best decision threshold adds no extra load when running the system because it is set ahead of time while testing the model. Instead of calculating on the fly, the chosen cutoff remains constant once the model goes live. During processing time, each transaction requires checking whether its fraud score exceeds the preset level. The check occurs quickly enough to fit within normal payment workflows. Users who complete valid transactions do not experience noticeable wait times. The performance remains smooth because the logic fits neatly into the existing steps. Detection runs in step with processing, not apart from it.

Instead of relying on heavy deep learning during transactions, the system uses a flexible design that fits fast-paced online banking environments. Because fraud detection works instantly, it can trigger quick responses, such as halting payments or asking for extra checks. This approach reduces monetary exposure without disrupting user interaction with the service.

\subsection{Supporting Phishing Detection Implementation}
A small extra layer of protection comes from a Chrome add-on that checks website addresses for signs of phishing attacks. This tool monitors every page visit and sends links to a server for review. Risk levels are judged using smart algorithms and pattern rules. Although it runs apart from the main fraud-tracking system, its presence helps guard against online scams. Protection improves when threats from harmful websites are detected early.

\subsection{Implementation Details}
A design built in separate parts allows for growth and simplifies updates. Prediction tools for spotting fraud run inside a server layer, offering real-time results via accessible endpoints. Communication between bank systems and this layer occurs through protected API requests, linking transactions and analysis smoothly. Speed, consistency across runs, and a straightforward setup guide how the system fits into active banking platforms.

\begin{table}[!t]
\centering
\caption{Performance Comparison with Baseline Models}
\resizebox{\columnwidth}{!}{
\begin{tabular}{lccccc}
\toprule
Model & Accuracy (\%) & Precision & Recall & F1-Score & ROC-AUC \\
\midrule
Logistic Regression & 95.4 & 82.6 & 71.2 & 76.5 & 0.942 \\
Random Forest & 97.1 & 90.3 & 88.7 & 89.5 & 0.981 \\
XGBoost (Fixed Threshold $\tau = 0.5$) & 97.8 & 94.1 & 92.4 & 93.2 & 0.988 \\
Proposed Model (Optimized $\tau$) & 98.3 & 96.8 & 98.6 & 98.2 & 0.996 \\
\bottomrule
\end{tabular}
}
\end{table}

\begin{figure}[t]
\centering
\includegraphics[width=0.7\columnwidth]{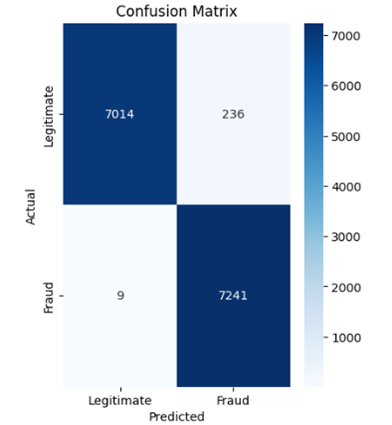}
\caption{Confusion matrix of the proposed fraud detection model on the test dataset.}
\end{figure}

\begin{figure}[!b]
\centering
\includegraphics[width=0.7\columnwidth]{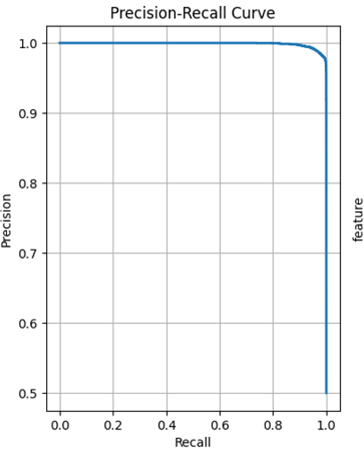}
\caption{illustrates the Precision–Recall (PR) curve of the proposed fraud detection framework under highly imbalanced class conditions. The curve demonstrates consistently high precision across a wide range of recall values, indicating that the model effectively identifies fraudulent transactions while minimizing false positives. This behavior is particularly important in online banking environments, where excessive false alarms lead to operational overhead. The strong PR performance confirms that the proposed cost-sensitive threshold optimization successfully prioritizes fraud recall without sacrificing precision.}
\end{figure}

\begin{figure}[t]
\centering
\includegraphics[width=0.8\columnwidth]{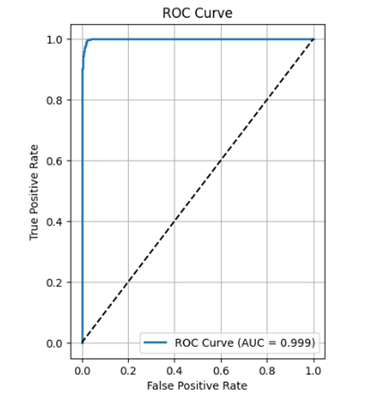}
\caption{presents the Receiver Operating Characteristic (ROC) curve of the proposed fraud detection model. The curve exhibits a steep ascent toward the upper-left corner, indicating strong discriminative capability between fraudulent and legitimate transactions. While the high ROC-AUC value reflects effective class separability, Precision–Recall analysis is emphasized in this work as it provides a more representative evaluation metric for fraud detection tasks characterized by severe class imbalance.}
\end{figure}

\begin{figure}[t]
\centering
\includegraphics[
  width=0.9\columnwidth,
  height=0.28\textheight,
  keepaspectratio
]{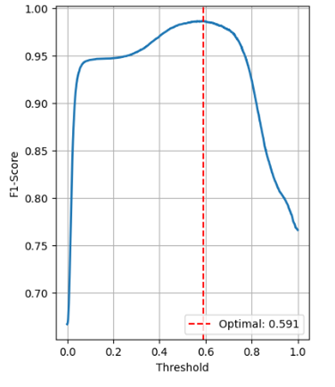}
\caption{Effect of decision threshold on model performance, illustrating the trade-off between detection accuracy and classification sensitivity.}
\end{figure}

\FloatBarrier
\section{CONCLUSION AND FUTURE WORK}
A closer look at fraud in online banking reveals how crucial it is to catch every suspicious case; missing one could have serious consequences. Instead of merely cutting costs, the method presented in this study focuses on reducing missed detections through smart modeling choices. Ensemble techniques are used to handle data where fraud cases are rare by using multiple models together while carefully adjusting decision boundaries. The results show that it works well - not perfectly, but solidly--when tested across different metrics, such as recall and classification accuracy. Visual tools such as ROC curves and confusion matrices confirmed the model’s ability to effectively separate normal transactions from fraudulent ones. What stands out is how smoothly the model fits into live transaction flows, proving that it can run fast enough for real-world use without delays. Beyond direct fraud spotting, another layer checks incoming links for signs of phishing attacks before they lead users to be deceived. This extra step does not replace the main system; rather, it supports it by closing gaps that pure transaction monitoring might leave open.

The results show that cost-aware decisions are important when spotting fraud in online banks. Instead of just measuring how often predictions are correct, aiming for high recall fits the actual risks that money institutions face. Threshold tuning plays a key role in ensuring that such systems function well under pressure.

\subsection{Future Work}
Looking ahead, efforts will be aimed at adapting the framework for real-world imbalances in class distribution alongside shifting fraud behaviors. Instead of fixed rules, thresholds can shift with transaction risk levels, guided by the practical system limits. Behavior over time may be modeled to reflect how users act over extended periods. Updating models step-by-step using continuous learning methods may offer a path forward. To test the wider applicability, broader transaction datasets must be considered. Real-system testing at scale is expected to follow, ensuring stability beyond controlled settings.

\end{document}